\documentclass[aps,eqsecnum,twocolumn,pacs]{revtex4}
\usepackage{epsfig}
\usepackage{graphicx}
\usepackage{amsmath}
\usepackage{amssymb}
\usepackage{color}
\newcommand{\be}{\begin{equation}}
\newcommand{\ee}{\end{equation}}
\def\ba{\begin{eqnarray}}
\def\ea{\end{eqnarray}}

\begin{document}
\title[Magnetic field driven instability of charged center in graphene]%
{Magnetic field driven instability of charged center in graphene}%
\author{O. V. Gamayun, E. V. Gorbar, and V. P. Gusynin}
\affiliation{Bogolyubov Institute for Theoretical Physics, 03680 Kiev, Ukraine}%

\begin{abstract}
It is shown that a magnetic field dramatically affects the problem of
supercritical charge in graphene making any charge in gapless theory
supercritical. The cases of radially symmetric potential well and
Coulomb center in an homogeneous magnetic field are considered. The local
density of states and polarization charge density are calculated in the first
order of perturbation theory. It is argued that the magnetically
induced instability of the supercritical Coulomb center  can be considered
as a quantum mechanical counterpart of the magnetic
catalysis phenomenon in graphene.
\end{abstract}
\maketitle

\section{Introduction}

Recently it was shown \cite{Shytov,Pereira,Novikov,Terekhov}  that atomic collapse
in a strong Coulomb field \cite{Greiner, Zeldovich}, a fundamental quantum
relativistic phenomenon still inaccessible in high-energy experiments, can be
readily investigated in graphene. In quantum electrodynamics (QED), taking into
account the finite size of nucleus \cite{finite-size}, theoretical works on the
Dirac-Kepler problem showed that for atoms with nuclear charge in excess of $Z
> 170$ the electron states dive into the lower continuum leading to positron
emission \cite{Greiner, Zeldovich}.

In graphene, the effective Coulomb coupling constant is given by $\beta=Z\alpha/\kappa$,
where $\alpha=e^2/\hbar v_F\simeq2.19$ is the ``fine-structure'' coupling constant,
$v_F \approx 10^6 m/s$ is the velocity of Dirac quasiparticles, and $\kappa$ is
a dielectric constant. The Hamiltonian of the system is not
self-adjoint when the coupling $\beta$ exceeds the critical value $\beta_{c}=1/2$ [1-4].
Similar to  the Dirac equation in QED,
one should replace the singular $1/ r$ potential by a regularized potential which
takes into account the finite size of the charged impurity, $R$: $V(r)=-\frac{Ze^{2}}{\kappa r}
\theta(r-R)-\frac{Ze^{2}}{\kappa R}\theta(R-r)$.  For gapped quasiparticles in such a
regularized potential, the critical coupling is determined by
$\beta_{c}=1/2+\pi^2/\log^2(c\Delta R/\hbar v_{F})$
\cite{excitonic-instability}, where $\Delta$ is a quasiparticle gap and the
constant $c\approx 0.21$, and $\beta_{c}$ tends to $1/2$ for $\Delta\to0$ or $R\to0$.

Since the electrons and holes strongly interact by means of the Coulomb interaction, one
may expect \cite{excitonic-instability,Sabio,Fertig} an excitonic instability
in graphene with subsequent phase transition to a phase with gapped
quasiparticles that may turn graphene into an insulator. This
semimetal-insulator transition in graphene is actively studied in the
literature, where numerical simulations give the critical coupling constant
$\alpha_c \approx 1.19$ \cite{Drut,Dillenschneider}.

In a many body system or quantum field theory, the  supercritical
coupling leads to more dramatic consequences compared to the case of
the Dirac equation for the electron in the Coulomb potential. Unlike the case
of the Coulomb center, the many body supercritical coupling instability cannot
be resolved through a spontaneous creation of a finite number of
electron-positron pairs. Like the Cooper instability in the theory of
superconductivity, the QED supercritical coupling instability is resolved only
through the formation of a condensate of electron-positron pairs generating a
mass gap in the spectrum \cite{Rivista}.

The presence of a magnetic field makes the situation even more interesting.
It was shown in \cite{GMSh1} that magnetic field catalyses the gap generation
for gapless fermions in relativistic-like systems and even the weakest attraction
leads to the formation of a symmetry breaking condensate. Therefore, the system
is always in the supercritical regime once there is an attractive interaction.
The magnetic catalysis plays an important role in quantum Hall effect studies in
graphene
\cite{Khveshchenko,graphite,MC-graphene-1,MC-graphene-2,MC-graphene-3,MC-graphene-4},
where it is responsible for lifting the degeneracy of the Landau levels.

The magnetic
catalysis phenomenon suggests that the Coulomb impurity in a magnetic field in
graphene should be supercritical for any $Z$. The Dirac equation for
quasiparticles in graphene in the Coulomb potential in a magnetic field was
considered in \cite{Khalilov-attempt} where exact solutions were found for
certain values of magnetic field, however, no instability or
resonance was found.

In QED in (3+1) dimensions, the Coulomb center problem in a magnetic field was
studied in \cite{Oraevskii}. There it was found that magnetic field $B$
confines the transverse electronic motion and the electron in the magnetic
field is closer to the nucleus than in the free atom. Thus, it feels stronger
Coulomb field. Therefore, $Z_c\alpha$ decreases with $B$. This result is
consistent with the magnetic catalysis phenomenon \cite{GMSh1}, according to
which, magnetic field catalyses gap generation and leads to zero critical
coupling constant in both (3+1)- and (2+1)-dimensional theories.

We would like to stress that the presence of an homogeneous magnetic
field changes qualitatively the supercritical Coulomb center problem. Indeed,
if magnetic field is absent, then the supercritical Coulomb center instability
leads to a resonance which describes an outgoing positron propagating freely to
infinity. However, since charged particles confined to a plane do
not propagate freely to infinity in a magnetic field, such a behavior is
impossible for the in-plane Coulomb center problem in a magnetic field.
Therefore, a priori it is not clear how the instability suggested by the
magnetic catalysis manifests itself in the Coulomb center problem in a magnetic
field. To answer this question is the main aim of this paper.

In Secs.~\ref{secII}, \ref{secIII} we consider the Dirac equation for
the electron in the potential well and Coulomb center in a magnetic field. We
study the local density of states (LDOS) and induced charge density for both
cases in Sec.~\ref{secIV}, where similarities and differences between the cases
of gapped and gapless quasiparticles as well as the potential well and Coulomb
interactions are discussed. In Sec.~\ref{SecV} we give a brief summary of our
results. Finally, we provide the details of our calculations of the LDOS and
polarization charge density in Appendix A.

\section{Potential well}
\label{secII}

The electron quasiparticle states in vicinity of the $K_{\pm}$ points of
graphene in the field of Coulomb impurity and in an homogeneous
magnetic field perpendicular to the plane of graphene are described by the
Dirac Hamiltonian in 2+1 dimensions
\be
H=\hbar
v_{F}\boldsymbol\tau\boldsymbol{p}+\xi\Delta\tau_{3}+V(r),
\ee
where the canonical momentum $\boldsymbol{p}=-i\boldsymbol\nabla +e\boldsymbol{A}/c$
includes the vector potential $\boldsymbol{A}$ corresponding to the external
magnetic field $\boldsymbol{B}$, $\tau_{i}$ are the Pauli matrices, and
$\Delta$ is a quasiparticle gap. The two component spinor $\Psi_{\xi s}$
carries the valley ($\xi=\pm)$ and spin ($s=\pm$) indices. We will use the
standard convention: $\Psi^{T}_{+s}=(\psi_{A},\psi_{B})_{K_{+}s}$ whereas
$\Psi^{T}_{-s}=(\psi_{B},\psi_{A})_{K_{-}s}$, and $A,B$ refer to two
sublattices of hexagonal graphene lattice. Since the interaction
$V(r)$ does not depend on spin, in what follows we will omit the spin index
$s$.

It is instructive to consider first the Dirac equation for the electron
in a potential well $V(r)=-V_0\theta(r_0-r)$ with $V_0
> 0$ in a magnetic field perpendicular to the plane of graphene.
We have
\[
\left(\begin{array}{cc} \xi\Delta &\hbar v_{F}(-iD_{x}-D_{y})\\
\hbar v_{F}(-iD_{x}+D_{y})&-\xi\Delta
\end{array}\right)\Psi(\mathbf{r})
\]
\be =\left(E-V(r)\right)\Psi(\mathbf{r}), \ee where $D_i=\partial_i+(ie/\hbar
c)A_i$ with $i=x,y$ is the covariant derivative and the symmetric gauge
$(A_{x},A_{y})=(B/2)(-y,x)$ is used for the magnetic field. It is clear that
the solution at $K_{-}$ point is obtained from the solution at $K_{+}$ point
changing $\Delta\to-\Delta$ and exchanging the spinor components
$\psi_{A}\leftrightarrow\psi_{B}$.

In polar coordinates
\ba
iD_{x}+D_{y}&=& e^{-i\phi}\left(i\frac{\partial}{\partial r}+\frac{1}{r}
\frac{\partial}{\partial\phi} +\frac{ieBr}{2\hbar c}\right),\nonumber\\
iD_{x}-D_{y}&=&e^{i\phi}\left(i\frac{\partial}{\partial r}-
\frac{1}{r}\frac{\partial}{\partial\phi} -\frac{ieBr}{2\hbar c}\right).
\ea
We can represent $\Psi(\mathbf{r})$ in terms of the eigenfunctions of the
conserved angular momentum $J_{z}=L_{z}+\sigma_{z}/2 =
-i\partial/\partial\phi+\sigma_{z}/2$ as follows:
\ba
\Psi(\mathbf{r})=\frac{1}{r}\left(\begin{array}{c} e^{i(j-\frac{1}{2})\phi}f(r)\\
ie^{i(j+\frac{1}{2})\phi}g(r)\end{array}\right), \label{psi-through-f-g} \ea
with $j=\pm 1/2, \pm 3/2, \dots$. For functions $f(r),\,g(r)$ we get the
following equations: \be
f^{\prime}-\frac{j+1/2}{r}f-\frac{r}{2l^{2}}f+\frac{E+\xi\Delta-V(r)}{\hbar
v_{F}}g=0, \label{f-eq}
\ee
\be
g^{\prime}+\frac{j-1/2}{r}g+\frac{r}{2l^{2}}g-\frac{E-\xi\Delta-V(r)}{\hbar
v_{F}}f=0, \label{g-eq}
\ee
where $l=\sqrt{\hbar c/|eB|}$ is the magnetic
length. These equations are easily solved for the potential well in
two regions $r<r_0$ and $r>r_0$ in terms of confluent hypergeometric functions.
In the region $r<r_0$, eliminating the function $g(r)$, we obtain the second
order differential equation for the function $f(r)$:
\be
f^{\prime\prime}-\frac{1}{\rho}f^{\prime}+\hspace{-0.5mm}\left[2p_V^2-j-\frac{1}{2}
-
\frac{j^{2}-j-3/4}{\rho^{2}} -\frac{\rho^{2}}{4}\hspace{-0.5mm}\right]\hspace{-0.5mm}f=0,\\
\label{f-Dirac-equation}
\ee
and in the region $r>r_0$ we have the same equation but with $V_{0}=0$.
Here we introduced the following dimensionless quantities:
\ba
p_V^2=\frac{l^{2}[(E+V_{0})^{2}-\Delta^{2}]}{2(\hbar v_{F})^{2}},\,
p^2=\frac{l^{2}(E^{2}-\Delta^{2})}{2(\hbar v_{F})^{2}},
\ea
and $\rho=r/l$. We get the following solutions which are regular at $r=0$,
\ba
f_{1}(\rho)&=&\rho^{j+\frac{1}{2}}e^{-\rho^{2}/4}\frac{C_{1}}{\Gamma(j+1/2)}\nonumber\\
&\times&\Phi\left(j+\frac{1}{2}
-p^{2}_{V},
j+\frac{1}{2};\frac{\rho^{2}}{2}\right),\\
g_{1}(\rho)&=&\frac{l(E+V_{0}-\xi\Delta)}{\sqrt{2}\hbar v_{F}}
\rho^{j+\frac{3}{2}}e^{-\rho^{2}/4}\frac{C_{1}}{\Gamma(j+3/2)}\nonumber\\
&\times&\Phi\left(j+\frac{1}{2}-p^{2}_{V},j+\frac{3}{2};\frac{\rho^{2}}{2}\right),
\ea
and decrease at infinity,
\ba
f_{2}(\rho)&=&C_{2}\rho^{j+\frac{1}{2}}e^{-\rho^{2}/4}\nonumber\\
&\times&\Psi\left(j+\frac{1}{2}
-p^{2},j+\frac{1}{2};\frac{\rho^{2}}{2}\right),\\
g_{2}(\rho)&=&\frac{\sqrt{2}\hbar v_{F}C_{2}}{l(E+\xi \Delta)}\rho^{j+\frac{3}{2}}e^{-\rho^{2}/4}\nonumber\\
&\times&\Psi\left(j+\frac{1}{2} -p^{2},
j+\frac{3}{2};\frac{\rho^{2}}{2}\right),
\ea
respectively [note that these expressions are valid at all $j=\pm 1/2,\pm 3/2,\dots$].

Then sewing solutions at $r=r_0$,
\be
\frac{f_{1}(\rho)}{f_{2}(\rho)}\Big|_{\rho=\rho_{0}}=\quad\frac{g_{1}(\rho)}{g_{2}(\rho)}\Big|_{\rho=\rho_{0}},
\quad \rho_{0}=\frac{r_{0}}{l}
\ee
we obtain the following transcendental equation for energies
of solutions with the total angular momentum $j$:
\ba
&&\frac{2(\hbar v_{F})^{2}(j+\frac{1}{2})\Phi\left(j+\frac{1}{2}-p^{2}_{V},j+\frac{1}{2};\frac{\rho^{2}_{0}}
{2}\right)}{l^{2}(E+V_0
-\xi\Delta)\Phi\left(j+\frac{1}{2}-p^{2}_{V},j+\frac{3}{2};\frac{\rho^{2}_{0}}
{2}\right)}\nonumber\\
&&=(E+\xi\Delta)\frac{\Psi\left(j+\frac{1}{2}-p^{2},j+\frac{1}{2};\frac{\rho^{2}_{0}}
{2}\right)}{\Psi\left(j+\frac{1}{2}-p^{2},j+\frac{3}{2};\frac{\rho^{2}_{0}}{2}\right)}\,.
\label{sewing}
\ea
Below we analyze this equation analytically and numerically.

\subsection{Instability in the absence of magnetic field}

In this subsection, we consider  the problem of the potential well instability
in the absence of magnetic field ($B=0$) which will serve us as a useful reference
point in the next section where we study instability in a magnetic field.
For $B=0$, the energy spectrum may be obtained either solving the Dirac
equation from the very beginning, or taking the limit of zero field
($l\to\infty$) in Eq.(\ref{sewing}). In the last case one needs to use the
following formulas \cite{Erdelyi},
\ba
\Phi(a,c;z)=e^{z}\Phi(c-a,c;-z),\\
\lim_{a\to\infty}\Phi(a,c;- z/a)=\Gamma(c)z^{\frac{1-c}{2}}J_{c-1}(2\sqrt{z}),
\\
\lim_{a\to\infty}[\Gamma(1+a-c)\Psi(a,c;-z/a)]\nonumber\\
=-i\pi e^{i\pi c}z^{\frac{1-c}{2}}H^{(1)}_{c-1}(2\sqrt{z}),\quad {\rm Im}z>0,
\ea
and $|{\rm arg}\,a|<\pi$ for the last two equations.

Assuming $|E|<\Delta$ we obtain
\ba
&&\frac{\sqrt{(E+V_{0})^{2}-\Delta^{2}}}{E+V_{0}-\xi \Delta}\frac{J_{j-1/2}(\beta r_{0})}
{J_{j+1/2}(\beta r_{0})}\nonumber\\
&&=\frac{\sqrt{E^{2}-\Delta^{2}}}{E-\xi \Delta}\frac{H^{(1)}_{j-1/2}(\beta'
r_{0})} {H^{(1)}_{j+1/2}(\beta' r_{0})}, \label{spectrum-Bzero}
\ea
where
$J_{\nu}(z)$, $H^{(1)}_{\nu}(z)$ are the Bessel and Hankel functions,
respectively, $\beta=\sqrt{(E+V_{0})^{2}-\Delta^{2}}/\hbar v_{F}$,
$\beta'=\sqrt{E^{2}-\Delta^{2}}/\hbar v_{F}$, and the square roots
are defined as ${\rm Im}\beta, {\rm Im}\beta'>0$. In the regions with ${\rm
Im}\beta, {\rm Im}\beta'\neq0$ one can use the relations $H^{(1)}_{\nu}(i
z)=(2/\pi i) e^{-i\pi\nu/2}K_{\nu}(z), J_{\nu}(i z)=e^{i\pi\nu/2}I_{\nu}(z)$.
 Eq.(\ref{spectrum-Bzero}) is invariant under the change ${j\to-j,\xi\to-\xi}$
\cite{footnote}.

Taking for the definiteness the $K_{-}$ point ($\xi=-$), one can see from
Eq.(\ref{spectrum-Bzero}) that the energy spectrum is continuous for
$|E|>\Delta$ and a discrete one for $|E|<\Delta$. The first bound state
$E\lesssim\Delta$ appears at an arbitrary small interaction $V_{0}$. Indeed,
taking $j=-1/2$ that corresponds to the smallest centrifugal barrier, we find
\be
E\simeq\Delta\left[1-2\left(\frac{\hbar v_{F}}{\Delta r_{0}}\right)^{2}
\exp\left(-\frac{2(\hbar v_{F})^{2}}{V_{0}\Delta r_{0}}-2\gamma\right)\right],
\label{weakboundstate}
\ee
where $\gamma$ is the Euler constant. Note that
there is no solution with the energy $E\lesssim\Delta$ at the $K_{-}$ point
with angular momentum $j=1/2$, but   such a solution exists at the $K_{+}$ point
similarly to the case of the Coulomb potential
\cite{excitonic-instability}.

As $V_0$ grows, at the critical strength of interaction
\be
V_{0\,cr}=\Delta\left[1+\sqrt{1+\left(\frac{\hbar v_{F}}{\Delta
r_{0}}\right)^{2}j^{2}_{0,1}}\right],
\label{V0crit}
\ee
($j_{0,1}\approx2.41$
is the first zero of the Bessel function $J_{0}(x)$) the lowest in energy bound
state dives into the lower continuum ($E=-\Delta$). We note that for the zero
gap case ($\Delta=0$) there are no bound states at all. In the
supercritical regime for $V_{0}>V_{0 cr}= \hbar v_{F}j_{0,1}/r_{0}$ (which
follows from Eq.(\ref{V0crit}) at $\Delta=0$) resonances with complex energies
appear leading to instability of the system similar to the case of the
supercritical Coulomb center \cite{Shytov,excitonic-instability}. The
occurrence of resonant states synchronously with diving into the lower continuum
of the lowest in energy bound state is the standard characteristic of QED
systems \cite{Shytov,Greiner,Zeldovich}. We will see in the next subsection
that the presence of an homogeneous magnetic field changes this conclusion.

Near the critical value of coupling the energy of resonant state is
given by
\ba
&&E=-\frac{V_{0}-V_{0 cr}}{\ln(1/ \delta)}\exp\left({
\frac{i\pi}{2\ln(1/ \delta)}}\right),
\nonumber\\
&&\delta = \frac{(V_{0}-V_{0 cr})r_0e^{\gamma-1}}{2\hbar v_F}, \quad 0<\delta\ll
1.
\ea
The dependence of energy on the  $V_{0}-V_{0 cr}$ (deviation
from the critical value) for the potential well interaction is nonanalytical
one but differs from the essential singularity that takes place in the Coulomb
center problem. This, of course, is related to the absence of scale invariance
for the potential well $V(r)$.

\subsection{Magnetically driven instability}

Before we consider the instability of the potential well problem in an
homogeneous magnetic field it is useful to recall the Landau energy levels for
the electron states in graphene in a magnetic field. If the interaction
vanishes ($V_{0}=0,r_{0}\to0$), Eq.(\ref{sewing}) gives the well known spectrum
of Landau levels:
\be
E=-\xi\Delta,\quad j\le-\frac{1}{2}, \label{LLL}
\ee
\be
E=\pm\sqrt{\Delta^{2}+2n\left(\frac{\hbar v_{F}}{l}\right)^{2}},\,n=1,2,\dots,\,
j+\frac{1}{2}\le n. \label{LL-1}
\ee
Note that the level $E=\Delta$ ($E=-\Delta$) is present only at the $K_{-}$ ($K_{+}$) point.

\begin{figure}[ht]
\includegraphics[width=\linewidth]{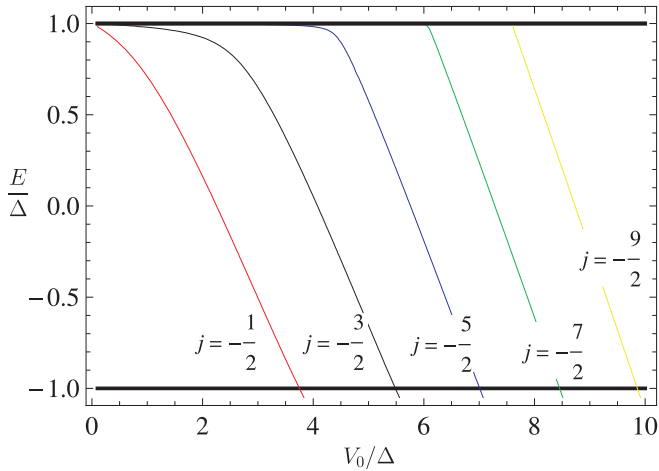}
\caption{The evolution of degenerate solutions of the lowest Landau
level at the $K_{-}$ point as a function of the dimensionless ratio $V_{0}/\Delta$.}
\label{Fig1}
\end{figure}

For nonzero $V_0$, the Landau energy levels are no longer
degenerate. Using the sewing equation (\ref{sewing}), we can determine the
evolution of degenerate solutions with $V_0$. For solutions of the Landau level
$E=\Delta$ with different $j$, their energies as function of $V_0/\Delta$ (at
fixed magnetic field $B$) are plotted in Fig.~\ref{Fig1} for $l\Delta/(\sqrt{2}\hbar v_{F})
=0.1$ and $\rho_0=r_0/l=0.02$.
We see that as $V_0$ increases more and more solutions with different $j$ {\it
cross the energy level $E=-\Delta$.} As usual \cite{Zeldovich,Greiner},
this means that vacuum of the second quantized theory is unstable with respect
to the creation of electron-hole pairs. However, as we discussed in
Introduction, there are {\it no resonance solutions} in the presence of constant
magnetic field. The reason for that is the presence of the positive $r^2/4l^2$
term due to the magnetic field in the effective Schr$\ddot{o}$dinger-like
equation for one component of the spinor function (see Eq.(\ref{U-1}) and Fig.3
in the next section) which qualitatively changes the asymptotic of the
effective potential: in nonzero field it grows at infinity instead of
decreasing as in the case $B=0$. Therefore quasiparticles are confined in such
a potential and cannot escape to infinity forming only discrete levels.

We would like to note that the situation under consideration is
analogous to that for a deep level vacancy in a many electron atom. There
electron states as solutions of the Dirac equation in the Coulomb potential of
the nucleus are stable. However, taking into account the interaction with the
second quantized electromagnetic field, the electrons on higher energy levels
are unstable with respect to the transition to the vacant state with emission of
photons.

The critical potential $V_{0 cr}$ is defined as the potential for which the first
crossing occurs. According to Fig.1, such a crossing is first
realized for the state with $j=-1/2$ (the potential well interaction lifts the
degeneracy of the Landau levels in quantum number $j$). Let us analyze in
detail how this state evolves with $V_0$. For the state with $j=-1/2$
Eq.(\ref{sewing}) becomes
\ba
&&(E+V_{0}-\Delta)\frac{\rho^{2}_{0}}{2}\frac{\Phi\left(1-p^{2}_{V},2;
\frac{\rho^{2}_{0}}{2}\right)}{\Phi\left(-p^{2}_{V},1;
\frac{\rho^{2}_{0}}{2}\right)}\nonumber\\
&&=-(E-\Delta)\frac{\Psi\left(-p^{2},0;
\frac{\rho^{2}_{0}}{2}\right)}{\Psi\left(-p^{2},1;
\frac{\rho^{2}_{0}}{2}\right)}, \label{eq:critvalue}
\ea
where we used the relation
\ba
\lim_{c\to-m}\frac{\Phi(a,c;x)}{\Gamma(c)}=\frac{\Gamma(a+m+1)}{\Gamma(a)(m+1)!}x^{m+1}\nonumber\\
\times \Phi(a+m+1,m+2;x),\,m=0,1,\dots. \label{formula}
\ea
For $V_0\to 0$, Eq.(\ref{eq:critvalue}) implies the following bound state at the
$K_{-}$ point:
\be
E=\Delta-V_{0}\left(1-e^{-r^{2}_{0}/2l^{2}}\right),
\ee
that is in contrast with
nonanalytical behavior in the coupling $V_{0}$ in the absence of magnetic field
described by Eq.(\ref{weakboundstate}). [At the $K_{+}$ point a similar bound
state exists but with angular momentum $j=+1/2$.]

\begin{figure}[ht]
\includegraphics[width=\linewidth]{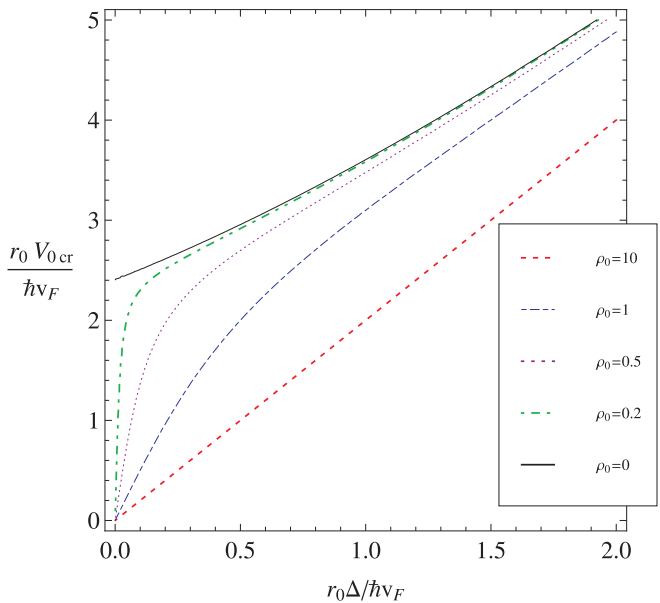}
\caption{The critical potential $V_{0cr}$ as a function of a gap for different
values of $\rho_0$. The case of zero magnetic field corresponds to
$\rho_{0}=0$.} \label{Fig2}
\end{figure}

As the coupling $V_{0}$ grows, energy of this bound state decreases
and finally crosses the level $E=-\Delta$ at some critical value $V_{0 cr}$. For
$E=-\Delta$, $p^{2}=0$, $p^{2}_{V}=l^{2}(V^{2}_{0}-2\Delta V_{0})/2({\hbar v_{F})^{2}}$, using
$\Psi(0,0;z)=\Psi(0,1;z)=1$ we find that Eq.(\ref{eq:critvalue}) defines the
following equation for $V_{0 cr}$:
\be
V_{0 cr}=2\Delta\left[1+\frac{2\Phi\left(a,1,\frac{\rho^{2}_{0}}{2}\right)}
{\rho_0^{2}\Phi\left(1+a,2,\frac{\rho^{2}_{0}}{2}\right)}\right],
\label{critical}
\ee
where $a=-l^{2}V_{0 cr}(V_{0 cr}-2\Delta)/2(\hbar v_{F})^{2}$. [Note that at
zero magnetic field ($l=\infty$) Eq.(\ref{critical})  reduces to Eq.({\ref{V0crit}})
for $V_{0 cr}$ that tends to a finite value in the gapless limit.] The critical potential
strength $V_{0cr}$  as a function of ${\Delta}$ is plotted in Fig.~\ref{Fig2} for
different values of the parameter $\rho_0$ which defines the ratio of
the potential well width to the magnetic length.
Analytically, it is not difficult to find that for $\rho_0 \ll 1$
Eq.(\ref{critical}) implies \be V_{0cr}=2{\Delta}(1+2l^2/r_0^2)\,.
\label{analytic}
\ee
It is clearly seen from Fig.\ref{Fig2} and from Eq.(\ref{analytic}) that
the critical potential strength
$V_{0cr}$ decreases with the growth of a magnetic field (or, with the decrease of $l$)
at fixed $r_{0}$ and $\Delta$. The physical reason for that is that the magnetic field
forces electron orbits to become closer to the charge center, making attraction
stronger and, thus, effectively lowering the critical coupling.

What is surprising here is that $V_{0cr}$ tends to zero as $\Delta \to 0$.
Thus, the presence of an homogeneous magnetic field leads to the instability of
gapless quasiparticles in the second quantized theory for {\it any} value of the
potential strength $V_0$. This result suggests that the Coulomb center in
gapless graphene in a magnetic field may be also unstable for any value $Ze$,
the problem which we study in the next section.

Finally, we will analyze states with energies near $\pm \Delta$ and large by modulus negative
momenta $j$. We find that there exists an infinite series of levels
approaching the energies $\pm\Delta$ asymptotically at large $|j+1/2|$ (i.e.,
for sufficiently large $j$ the effect of the potential interaction $V_0$ can be
neglected and the Landau levels are recovered). For $V_0 \to 0$, they behave as
\be
E\simeq -\xi\Delta-\frac{V_{0}e^{-\rho^{2}_{0}/2}}{\Gamma(k+1)}
\left(\frac{\rho^{2}_{0}}{2}\right)^{k+1},\quad
k=-(j+\frac{1}{2})>>1. \label{thresholdenergies}
\ee
This can be found directly by solving Eq.(\ref{sewing}), first taking there the
limit $j+1/2\to-k$ by means of Eq.(\ref{formula}) and analyzing then the equation
at weak coupling and large $k$. Alternatively, Eq.(\ref{thresholdenergies}) is
obtained as the first order correction in the interaction to the levels
$\pm\Delta$ at $K_{\pm}$ points in a magnetic field. Note that the levels
(\ref{thresholdenergies}) lie below $\Delta$ for the $K_{-}$ point and below
$-\Delta$ for the $K_{+}$ point, respectively.

\section{The Coulomb center}
\label{secIII}

The equations for the functions $f(r)$ and $g(r)$ for the Coulomb center
problem follow directly from Eqs.(\ref{f-eq}) and (\ref{g-eq}) by setting
$V(r)=-Ze^{2}/r\,\,\theta(r-R)-Ze^{2}/R\,\,\theta(R-r)$ there
(we take the dielectric constant $\kappa=1$).
Eliminating, for example the function $f(r)$, one can get a second order
differential equation for the function $g(r)$. Further, introducing the
function $\chi(r)$ by means of the relation
\be
[E-\xi\Delta-V(r)]^{1/2}\chi(r)=\frac{g(r)}{\sqrt{r}},
\ee
we get the Schr$\ddot{o}$dinger-like equation, \be -\chi''(r)+U(r)\chi(r)={\cal E}\chi(r),
\label{effective-equation} \ee where \be {\cal E}=E^{2}-\Delta^{2}, \ee and the
effective potential, $U=U_{1}+U_{2}$, \ba U_{1}&=&\frac{V(2E-V)}{(\hbar
v_{F})^{2}}+\frac{j(j+1)}{r^{2}}+\frac{r^{2}}{4l^{4}}
+\frac{j-1/2}{l^2},\label{U-1}\\
U_{2}&=&\frac{1}{2}\left[\frac{V''}{E-\xi\Delta-V}+\frac{3}{2}\left(\frac{V'}
{E-\xi\Delta-V}\right)^{2}\right.\nonumber\\
&-&\left.\left(\frac{j}{r}+\frac{r}{2l^{2}}\right)\frac{2V'}
{E-\xi\Delta-V}\right]. \label{U-2}
\ea
We plot effective potential $U(r)$
near $K_{-}$ point for $E=-\Delta$  and $j=-1/2$ in Fig. 3. There
the energy barrier in the absence of magnetic field is clearly seen, which
leads to the appearance of resonances for sufficiently large charge. The
presence of non-zero magnetic field {\it changes} the asymptotic of the
effective potential at infinity and, thus, forbids the occurrence of resonance
states. This feature distinguishes {\it qualitatively} the Coulomb center
problem (as well as the potential well problem) in a magnetic form from that at
$B=0$.
\begin{figure}[ht]\label{pot}
\includegraphics[width=\linewidth]{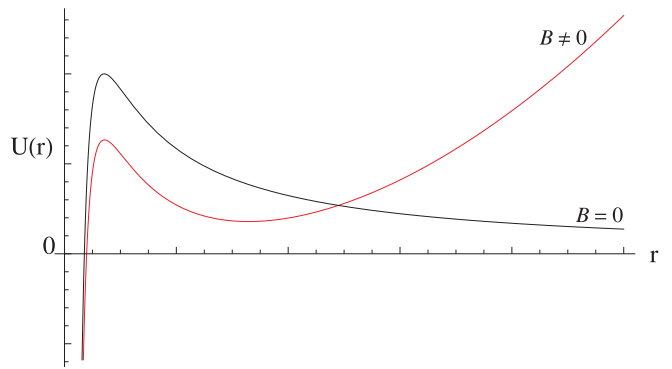}
\caption{The  potential $U(r)$ as a function of a distance from the Coulomb
center at zero and nonzero magnetic field for the state with $E=-\Delta$ and
$j=-1/2$.}
\end{figure}

Unfortunately, Eq.(\ref{effective-equation}) belongs to the class of
equations with two regular and one irregular singular (at $r=\infty$) points,
and exact solutions of this equation  cannot be expressed in closed form
in terms of  known special functions.

Since we are interested in solving Eqs.(\ref{f-eq}), (\ref{g-eq}) with the
Coulomb potential in the regime $Z\alpha \to 0$ ($\alpha=e^{2}/\hbar v_{F}$),
we can find solutions using perturbation theory. For $Z\alpha=0$, solutions of
Eqs.(\ref{f-eq}), (\ref{g-eq}) are the well known that
 Landau states degenerate in
the total angular momentum $j$. For the level $E^{(0)}=\Delta$ their normalized
wave functions (\ref{psi-through-f-g}) have the form (at the $K_{-}$ point)
\begin{equation}
\Psi_k(r,\phi)=\frac{(-1)^{k}}{l\sqrt{2\pi k!}}e^{-r^2/4l^{2}} \left( \begin{array}{c} 0 \\
\left(\frac{r^{2}}{2l^{2}}\right)^{k/2}e^{-i k\phi}
\end{array} \right),
\label{unperturbed-solution}
\end{equation}
 where $k=-(j+1/2)=0,1,2,...\,.$ The Coulomb potential removes degeneracy in
$j$. Energy corrections of perturbed states of the Landau level
$E^{(0)}=\Delta$ are found from the secular equation
$$
|E^{(1)}-V_{k_1k_2}|=0\,.
$$
Since $V_{k_1k_2}$ is a diagonal matrix, we easily obtain
\ba
E^{(1)}_k=V_{kk}&=&-\frac{Ze^{2}}{
k!2^{k}l}\int\limits_0^{\infty}d\rho\,\rho^{2k}\,e^{-\rho^2/2}\nonumber\\
&=&-\frac{Ze^{2}\Gamma(k+\frac{1}{2})}{l\sqrt{2}\Gamma(k+1)}\,.
\label{energy-correction}
\ea
Thus at large $k$ the energy levels accumulate
near $E=\Delta$:
\be E_k\simeq \Delta-\frac{Ze^{2}}{l\sqrt{2k}}.
\ee
Like in the case of the potential well interaction considered in the previous
section, the largest correction by modulus $E^{(1)}_0=-Z\alpha\hbar
v_F\sqrt{\pi}/l\sqrt{2}$ occurs for the state with $j=-1/2$ ($k=0$). The
critical charge is determined by the condition $E=E^{(0)}+E^{(1)}_{0}=-\Delta$
when the level $E$ crosses the level of filled state. This gives
\begin{equation}
Z_c\alpha=\frac{2\sqrt{2}\Delta l}{\sqrt{\pi}\hbar v_{F}}\,.
\label{critical-charge}
\end{equation}
Like in the case of the potential well in a magnetic field, the critical charge
(\ref{critical-charge}) tends to zero as $\Delta \to 0$. This means that
magnetic field indeed dramatically affects the Coulomb center problem in
graphene making any charge in gapless theory supercritical.
\begin{figure}[ht]
\includegraphics[width=\linewidth]{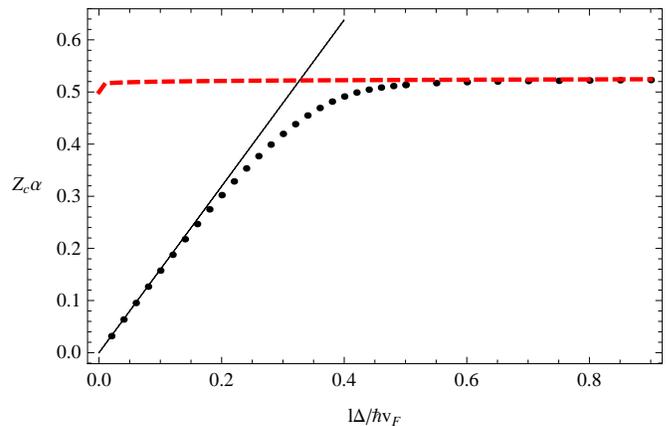}
\caption{The critical Coulomb coupling $Z_{c}\alpha$ as a function
of gap at zero (dashed red line) and nonzero magnetic field
(dotted black line) for the state with $j=-1/2$. The
straight black line corresponds to the critical Coulomb coupling in the first
order of perturbation theory given by Eq.(\ref{critical-charge}).}
\label{crit-Coulomb}
\end{figure}
Eq.(\ref{critical-charge}) gives the critical Coulomb coupling in the
regime $Z\alpha \to 0$ in the first order of perturbation theory. For arbitrary
values of $Z\alpha$, we calculated the dependence of the critical coupling on
the gap numerically. The corresponding results are presented in
Fig.~\ref{crit-Coulomb}, where, for the parameter regularizing the Coulomb
potential, we took $R=10^{-3}l$. The dashed red line in Fig.~\ref{crit-Coulomb}
gives the critical Coulomb coupling $Z_{c}\alpha=1/2+\pi^2/\log^2(c\Delta
R/\hbar v_{F})$ in the absence of magnetic field (see Fig. 1 in
Ref.\cite{excitonic-instability}). Thus, at weak magnetic field ($l\to\infty$)
the critical coupling tends to $1/2$ while $Z_{c}\alpha\to0$ for $l\Delta\to0$
in the gapless or strong magnetic field regime.

\section{The local density of states}
\label{secIV}

It is interesting to see how a magnetic field and the Coulomb center
affect the local density of states  of  quasiparticles that can be directly
measured in scanning tunneling microscope (STM) experiments. The crucial
difference of the case of gapless quasiparticles from that of gapped
ones in a magnetic field is that the critical charge is zero for gapless
quasiparticles, therefore, energies of all previously degenerate states of the
lowest Landau level become negative.

The LDOS at the distance $\mathbf{r}$ from impurity is given by
\be
\rho(E;\mathbf{r})=-\frac{1}{\pi}{\rm tr}{\rm Im}G(\mathbf{r},\mathbf{r};
E+i\eta),\quad \eta\to0,
\ee
where trace includes the summation over valley, sublattice and spin degrees of freedom.
The retarded Green's function $G(\mathbf{r},\mathbf{r'};E+i\eta)$ in a constant
magnetic field can be written in the form
\ba
&&G({\bf r},{\bf r}^{\prime}; E) =e^{i\Phi({\mathbf r},{\mathbf r}^\prime)}
\tilde{G}({\mathbf r},{\mathbf r}^\prime;E),\\
&& \Phi({\mathbf r},{\mathbf r}^\prime)= \frac{e}{\hbar c}\int \limits_{{\bf
r}^\prime}^{{\bf r}} A_i^{ext}(z)dz^i,
\ea
where we separated the gauge dependent (Schwinger) phase
 $\Phi({\mathbf r},{\mathbf r}^\prime)$ from a gauge invariant part of the Green's
 function $\tilde{G}({\mathbf r},{\mathbf r}^\prime;E)$ . The last one satisfies the
following Lippmann-Schwinger equation:
\ba
\tilde{G}({\bf r},{\bf r}^{\prime};
E) = \tilde{G}_{0}({\bf r}-{\bf r}^{\prime}; E)
+\int d{\bf r}^{\prime\prime}\tilde{G}_{0}({\bf r}-{\bf r}^{\prime\prime}; E)\nonumber\\
\times V({\bf r}^{\prime\prime})\tilde{G}({\bf r}^{\prime\prime},{\bf
r}^{\prime}; E) e^{i[\Phi({\bf r},{\bf r}^{\prime\prime})+\Phi({\bf
r}^{\prime\prime},{\bf r}^{\prime}) +\Phi({{\bf r}^{\prime},\bf r})]}.
\label{Lippmann-Schwinger1}
\ea
[Note that the Green function $\tilde{G}({\bf
r},{\bf r}^{\prime}; E)$ is not translation invariant in presence of an
impurity unlike the noninteracting function $\tilde{G}_{0}({\bf r}-{\bf
r}^{\prime}; E)$.] For weak interaction we can calculate the LDOS in the first
order in the perturbation theory,
\be
\rho(E;\mathbf{r})=\rho_{0}(E;\mathbf{r})+\delta\rho(E;\mathbf{r}),
\ee
where $\rho_{0}(E;\mathbf{r})$ is the LDOS for free quasiparticles in a magnetic
field, and
\be \delta\rho(E;\mathbf{r})=-\frac{1}{\pi}{\rm Im}\hspace{-1mm}\int
\hspace{-1mm} d{\bf r}^{\prime} {\rm tr}\hspace{-1mm}\left[\tilde{G}_{0}({\bf
r}-{\bf r}^{\prime})V({\bf r}^{\prime}) \tilde{G}_{0}({\bf r}^{\prime}-{\bf
r})\right].
\label{LDOS-correction-1storder}
\ee

First we consider the case of gapless quasiparticles. In this case,
$\rho_{0}(E;\mathbf{r})$ and $\delta\rho(E;\mathbf{r})$ are calculated for
gapless quasiparticles in Appendix. We got there that the LDOS decreases at
large distances ($r>>r_{0},l$) as
\ba
\delta\rho(\mathbf{r},E)\simeq\frac{V_{0}r^{2}_{0}}{2\pi(\hbar v_{F})^{2}} {\rm
Im}[\lambda\Gamma^{2}(-\lambda)]
e^{-\mathbf{r}^{2}/2l^{2}}\hspace{-1.5mm}\left(\frac{\mathbf{r}^{2}}{2l^{2}}
\right)^{2\lambda}
\label{induced-potential-well}
\ea for the potential well, while for the Coulomb center we obtained
\ba
\delta\rho(\mathbf{r},\mathbf{r},E)= \frac{Ze^{2}}{\kappa}\frac{1}{2\pi(\hbar
v_{F})^{2}}\left[\frac{B_{0}(\lambda)}{r}
+\frac{l^{2}B_{1}(\lambda)}{2r^{3}}\right]
\label{LDOS-correction-Coulomb}
\ea
with the functions $B_{i}(\lambda)$ given by Eqs.(\ref{function-B0}),
(\ref{function-B1}) and $\lambda$ is defined after
Eq.(\ref{confluent-vs-Whittaker}). For the induced charge density,
\be
n_{ind}(\mathbf{r})=-e\int\limits_{-\infty}^{0}dE\delta\rho(\mathbf{r},E),
\label{inducedcharge}
\ee
using Eqs.(\ref{induced-potential-well})
and (\ref{LDOS-correction-Coulomb}), we find that it is positive at large
distances and decreases exponentially for the potential well and, due to $\int_{-\infty}^{0}dE
B_{0}(\lambda)=0$, as $1/r^{3}$ for the Coulomb interaction,
\be
n_{ind}(\mathbf{r})\simeq a\frac{Ze^{3}l}{\kappa \hbar v_{F}}\frac{1}{r^{3}},
\quad a=-\frac{3\zeta(-1/2)}{2\pi\sqrt{2}}\approx0.07.
\ee
We remind that in the absence of magnetic field the polarization charge diminishes as
$1/r^{2}$ both in the supercritical $Z\alpha/\kappa>1/2$ Coulomb
center \cite{Shytov} and potential well $V_{0}>V_{0 cr}= \hbar
v_{F}j_{0,1}/r_{0}$ \cite{Milstein} interactions.

The situation is quite different in the case of gapped quasiparticles.
Here we will consider the polarization charge density for the most
interesting case of the Coulomb center in an homogeneous magnetic field. The
polarization charge density (\ref{inducedcharge}) can be rewritten in the more
familiar form
\be
n_{ind}(\mathbf{r})=-e\sum_{E
\le-\Delta}\left[\Big|\Psi_{E}(\mathbf{r})\Big|^{2}
-|\psi_{E}(\mathbf{r})|^{2}\right]\,,
\label{polarization-charge-density}
\ee
where $\psi_{E}$ and $\Psi_{E}$ are the Landau wave functions and the wave
functions of the Coulomb center problem in a magnetic field, respectively.
Since we consider the case where $Z\alpha$ is small, the corrections to the
wave functions of negative energy states of the deep Landau levels defined by
Eq.(\ref{LL-1})  can be ignored. We will consider only the
corrections to the lowest Landau level states given by Eq.(\ref{LLL}).

In the first order of perturbation theory, wave functions of the
Landau level $E=-\Delta$ are sought as superposition of all degenerate states
with unknown coefficients $\Psi_{E}=\sum_{j \le 1/2}c_j\psi_{-\Delta}^{j}$,
where $\psi_{-\Delta}^j$ are wave functions of the Landau level with
$E=-\Delta$ and total momentum $j$. The unknown $c_j$ are determined by the
equation
\be
\sum_{j_2\le-1/2}(V_{j_1j_2}-E^{(1)}\delta_{j_1j_2})\,c_{j_2}=0\,.
\label{perturbation-theory}
\ee
Since $V_{j_1j_2}=V_{j_1j_1}\delta_{j_1j_2}$ is
a diagonal matrix, the secular equation $|V_{j_1j_2}-E^{(1)}|=0$ is
trivially solved and $E_j^{(1)}=V_{jj}$. Then
Eq.(\ref{perturbation-theory}) implies that the corresponding unknown
coefficients $c_{j_1}$ are equal $c_{j_1}=1$ for $j_1=j$ and zero for all other
$j_1$. Notice that $c_j=1$ because the wave functions $\psi_{-\Delta}^j$ are
normalized. This means that the wave functions of perturbed states of the
Landau level $E=-\Delta$ do not change in the first order of perturbation
theory. Consequently, according to Eq.(\ref{polarization-charge-density}), they
do not contribute to the polarization charge density. Clearly, the polarization
charge appears only when the first perturbed state of the Landau level
$E=\Delta$ with $j=-1/2$ crosses the threshold of filled states of
the lowest Landau level $E=-\Delta$. Using
Eq.(\ref{unperturbed-solution}), (\ref{polarization-charge-density}), and the
fact that perturbed wave functions of the $E=\Delta$ Landau level states do not
change in the first order in the Coulomb potential, similarly to the case of
perturbed wave functions of the $E=-\Delta$ Landau level considered above, we
conclude that for the critical charge $Z_c\alpha$ given by
Eq.(\ref{critical-charge}) the polarization charge density equals
\be
n_{ind}(\mathbf{r})=-\frac{e}{2\pi\,l^2}e^{-r^2/2l^2}\,.
\label{polarization-charge-density-1}
\ee
Thus, the polarization charge density is concentrated near the impurity where
it is negative and quickly decreases at large distances.

\section{Conclusion}
\label{SecV}

In this paper we showed that in an external magnetic field the value of the
critical coupling for the onset of instability of a system of planar
Dirac gapless quasiparticles interacting with charged center (charged impurity)
reduces to zero. This result serves as a quantum mechanical single-particle
counterpart of the magnetic catalysis phenomenon in graphene. The cases of
radially symmetric potential well and Coulomb center were analytically
considered. The local density of states and induced charge density were
calculated in the first order of perturbation theory for gapless
quasiparticles.

The crucial ingredient for the instability is the existence of zero energy
level for gapless Dirac fermions in a magnetic field which is infinitely degenerate. In
this case any weak attraction leads to the appearance of empty states in the
Dirac sea of negative energy states and to the instability of a system.

One should stress a qualitative difference in the phenomenon of instability
between gapped and gapless quasiparticles. In the case of gapped quasiparticles
there is a {\it finite} critical value for the strength of interaction when the
lowest unfilled level crosses the first filled one forming a hole in the sea of
filled states. As the coupling grows, more and more levels cross that
level. Clearly, the system tries to rearrange itself filling in empty states
whose presence is a signal of instability. The important difference of the case
of gapless quasiparticles from that of gapped ones, besides the critical
coupling being zero, is that an infinite number of states of the previously
degenerate lowest Landau level become vacant.

Thus, the presence of an external magnetic field changes dramatically
the problem of atomic collapse in graphene in a strong Coulomb field
\cite{Shytov}. Clearly the problem becomes a many body one and requires field
theoretical methods to find a true ground state. One should expect that the gap
generation for initially gapless quasiparticles should take place already in
the weak coupling regime in the presence of a magnetic field \cite{graphite}.
\vspace{2mm}

\centerline{\bf Acknowledgements}
\vspace{1mm}
We are grateful to  V.A. Miransky and I.A. Shovkovy for useful
discussions. This work  is supported partially by the SCOPES under Grant
No. IZ73Z0-128026 of the Swiss NSF, the Grant No. SIMTECH 246937 of the European
FP7 program, by the SFFR-RFBR Grant ``Application of string theory and field theory
methods to nonlinear phenomena in low dimensional systems",  and by the Program of
Fundamental Research of the Physics and Astronomy Division of the NAS of Ukraine.

\section{Appendix}
The Green's function of free quasiparticles in a magnetic field is
well known (see, for example, \cite{GMSh1,graphite}), and in the configuration
space it has the form of a series over the Landau levels (we consider the zero
gap case),
\ba
&&\tilde{G}_{0}(\mathbf{r};E)=\frac{1}{2\pi l^{2}}\,e^{-\frac{\mathbf{r}^{2}}
{4l^{2}}}\sum\limits_{n=0}^{\infty}\frac{1}{(E+i\eta)^{2}-M_{n}^{2}}\nonumber\\
&&\times\left[E\left[P_{-}L_{n}\left(\frac{\mathbf{r}^{2}}
{2l^{2}}\right)+P_{+}L_{n-1}\left(\frac{\mathbf{r}^{2}}
{2l^{2}}\right)\right]\right.\nonumber\\
&&+\left.i\hbar v_{F}\frac{\boldsymbol{\tau}\mathbf{r}}{l^{2}}L^{1}_{n-1}
\left(\frac{\mathbf{r}^{2}}{2l^{2}}\right)\right],
\label{Landau-levels-expansion}
\ea
where $M_{n}=(\hbar v_{F}/l)\sqrt{2n}$ are the energies of the Landau levels,
$P_{\pm}=(1\pm\tau_{3})/2$ being the projectors, $L_{n}^{\alpha}(z)$ the
generalized Laguerre polynomials (by definition $L_{n}(z)\equiv L^{0}_{n}(z)$
and $L_{-1}^{\alpha}(z)\equiv 0$), and the Pauli matrices $\tau_{1,2}$ act in
the sublattice space.
 The sum over the Landau levels can be explicitly performed by means of the formula
\be
\sum\limits_{n=0}^{\infty}\frac{L_{n}^{\alpha}(x)}{n+b}=\Gamma(b)\Psi(b;1+\alpha;x)
\ee
(see, Eq.(6.12.3) in the book \cite{Erdelyi}), leading to a closed expression for the free
Green's function (see, recent papers \cite{Pyatkovskiy,Rusin}),
\ba
\tilde{G}_{0}(\mathbf{r};E)&=&-\frac{e^{-\frac{\mathbf{r}^{2}}
{4l^{2}}}}{4\pi \hbar^{2}v^{2}_{F}}\left\{E\left[P_{-}\Gamma(-\lambda)\Psi
\left(-\lambda;1;\frac{\mathbf{r}^{2}}{2l^{2}}\right)\right.\right.\nonumber\\
&+&\left.\left.P_{+}\Gamma(1-\lambda)
\Psi\left(1-\lambda;1;\frac{\mathbf{r}^{2}}{2l^{2}}\right)\right]\right.\nonumber\\
&+&\left.i\hbar
v_{F}\frac{\boldsymbol{\tau}\mathbf{r}}{l^{2}}\Gamma(1-\lambda)
\Psi\left(1-\lambda;2;\frac{\mathbf{r}^{2}} {2l^{2}}\right)\right\}.
\label{GFfree-final}
\ea
Here $\Psi(a;c;x)$ is the confluent hypergeometric
function which is related to the Whittaker function,
\be
\Psi(a;c;x)=e^{\frac{x}{2}}x^{-\frac{1}{2}-\mu}W_{\kappa,\mu}(x), \,
\kappa=\frac{c}{2}-a, \mu=\frac{c-1}{2}, \label{confluent-vs-Whittaker} \ee
$\Gamma(x)$ is the Euler gamma function, and $\lambda=(E+i\eta)^{2}
l^{2}/(2\hbar^{2} v_{F}^{2})$.

The LDOS of free quasiparticles in a magnetic field does not depend on
$\mathbf{r}$ and is given by
\ba
\rho_{0}(E)=-\frac{1}{\pi}\lim_{\mathbf{r}\to0}{\rm Im}{\rm
tr}[\tilde{G_{0}} (\mathbf{r};E+i\eta)]=
\frac{1}{\pi^{2} \hbar^{2}v^{2}_{F}}\nonumber\\
\times\lim_{\mathbf{r} \to 0}{\rm Im}\left\{(E+i\eta)\left[\Gamma(-\lambda)\Psi
\left(-\lambda;1;\frac{\mathbf{r}^{2}}{2l^{2}}\right)\right.\right.\nonumber\\
\left.\left.+\Gamma(1-\lambda)
\Psi\left(1-\lambda;1;\frac{\mathbf{r}^{2}}{2l^{2}}\right)\right]\right\}.
\ea
The hypergeometric function $\Psi(a;c;x)$ at small $x$ behaves as
\ba
\Psi(a;1;x)\simeq -\frac{1}{\Gamma(a)}[\ln x +\psi(a)+2\gamma]+O(x\ln x),\nonumber\\
\Psi(a;2;x)\simeq \frac{1}{\Gamma(a)x}+\frac{1}{\Gamma(a-1)}[\ln x +\psi(a)\nonumber\\
+2\gamma-1]+O(x\ln x),
\ea
where $\psi(z)$ is the digamma function. Therefore
\be \rho_{0}(E)=- \frac{1}{(\pi\hbar v_{F})^{2}}{\rm
Im}\left[(E+i\delta)\left(\psi(-\lambda) +\psi(1-\lambda)\right)\right],
\label{LDOS-free}
\ee
and the LDOS free quasiparticles in a magnetic field
finally is found to be
\be
\rho_{0}(E)=\frac{2}{\pi l^{2}}\left[\delta(E)+\sum\limits_{n=1}^{\infty}
[\delta(E-M_{n})+\delta(E+M_{n})]\right],
\ee
(compare with Eq.(4.2) in Ref.\cite{ShGB}).

The first order correction to the LDOS due to the interaction is given by
Eq.(\ref{LDOS-correction-1storder}). For the radial well to find the
asymptotic at distances $r\gg r_{0}$, where $r_{0}$ is a range of the
potential, we can put $r'=0$ in the arguments of the free Green's functions in
Eq.(\ref{LDOS-correction-1storder}) and get the following behavior: \ba
&&\delta\rho(\mathbf{r},\mathbf{r};E)=V_{0}r^{2}_{0}{\rm Im}{\rm tr}
[\tilde{G}_{0}({\bf r}; E)\tilde{G}_{0}(-{\bf r}; E)]\simeq\nonumber\\
&&\frac{2V_{0}r^{2}_{0}}{(\pi\hbar v_{F}l)^{2}}{\rm Im}[\lambda\psi\left(-\lambda\right)
]\ln\frac{\mathbf{r}^{2}}{2l^{2}},\\
&&\frac{V_{0}r^{2}_{0}}{2(\pi\hbar v_{F}l)^{2}}{\rm
Im}\left[\lambda\Gamma^{2}(-\lambda)\right]
e^{-\mathbf{r}^{2}/2l^{2}}\left(\frac{\mathbf{r}^{2}}{2l^{2}}\right)^{2|\lambda|}\hspace{-2mm},
\ea in the regions $l>>r>>r_{0}$ and $r>>\mbox{max}(l,r_{0})$, respectively. As
is seen, the correction to the free LDOS is an odd function of energy.

To find the first order correction due to the Coulomb potential we first calculate
the correction to the Green's function which is given by
\ba
&&\delta G(\mathbf{r},\mathbf{r};E)=-\frac{Ze^{2}}{\kappa}\int
d{\bf r}^{\prime}\tilde{G}_{0}({\bf r}-{\bf r}^{\prime}; E)
\frac{1}{|\mathbf{r}^{\prime}|}\tilde{G}_{0}({\bf r}^{\prime}-{\bf r}; E)\nonumber\\
&&=-\frac{Ze^{2}}{\kappa}\int d{\bf r}^{\prime}\tilde{G}_{0}({\bf r}^{\prime};
E) \frac{1}{|\mathbf{r}^{\prime}-\mathbf{r}|}\tilde{G}_{0}(-{\bf r}^{\prime};
E). \label{Coulombmagn-corr}
\ea
Taking trace over spin and Dirac indices,
performing integration over the angle by means of the formula
\be
\int\limits_{0}^{2\pi}\frac{d\theta}{\sqrt{r^{2}+{r^{\prime}}^{2}-2rr^{\prime}\cos\theta}}
=\frac{4}{r+r^{\prime}}
\,K\left(\sqrt{\frac{4rr^{\prime}}{(r+r^{\prime})^{2}}}\right),
\label{angle-integral}
\ee
where $K(k)$ is the elliptic integral of the first kind ($K(0)=\pi/2)$, and
calculating the imaginary part, we obtain
\ba
\hspace{-5mm}&&\delta\rho(\mathbf{r},\mathbf{r},E)=\frac{2Ze^{2}}{\kappa}
\frac{{\rm sgn}(E)}{(\pi\hbar
v_{F}l)^{2}}\int\limits_{0}^{\infty}\frac{dr^{\prime}r^{\prime}}
{r+r^{\prime}}e^{-x}\nonumber\\
\hspace{-5mm}&&K\left(\sqrt{\frac{4rr^{\prime}}{(r+r^{\prime})^{2}}}\right)
\left\{\sum\limits_{n=0}^{\infty}\left[\lambda\left([L_{n}(x)]^{2}+[L_{n-1}(x)]^{2}\right)
\right.\right.\nonumber\\
\hspace{-5mm}&&\left.\left.+2x[L^{1}_{n-1}(x)]^{2}\right]\delta'(\lambda-n)\right.\nonumber\\
\hspace{-5mm}&&\left.-\sum\limits_{n,m=0,n\neq m}^{\infty}\left[\lambda\left(L_{n}(x)L_{m}(x)
+L_{n-1}(x)L_{m-1}(x)\right)\right.\right.\nonumber\\
\hspace{-5mm}&&\left.\left.+2xL^{1}_{n-1}(x)L^{1}_{m-1}(x)\right]\frac{\delta(\lambda-n)-
\delta(\lambda-m)}{n-m}\right\},
\ea
where $x={r'}^{2}/2l^{2}$.
The correction to the LDOS at large distances $r>>l$ is given by Eq.(\ref{LDOS-correction-Coulomb})
where the energy dependence is given by the functions
\be
B_{0}(\lambda)={\rm sgn}(E)\left[\lambda\delta'(\lambda)+2\sum\limits_{n=1}^{\infty}(\lambda+n)
\delta'(\lambda-n)\right],\label{function-B0}\\
\ee
\ba
B_{1}(\lambda)&=&{\rm sgn}(E)\left[\lambda\delta'(\lambda)+4\sum\limits_{n=1}^{\infty}
n(\lambda+n)\delta'(\lambda-n)\right.\nonumber\\
&+&2\delta(\lambda-1)+2\left.\sum\limits_{n=1}^{\infty}[\lambda(2n+1)+2n(n+1)]\right.\nonumber\\
&\times&\left.\left(\delta(\lambda-n-1)
-\delta(\lambda-n)\right)\right].
 \label{function-B1}
\ea
To calculate the integrals of the Laguerre polynomials we used the
following generating function (see, Appendix A in \cite{Pyatkovskiy}):
\ba
\hspace{-6mm}&&I^{\alpha}_{nm}(y)=\int_{0}^{\infty}dt\,e^{-t}t^{\alpha}J_{0}(2\sqrt{yt})L_{n}^{\alpha}(t)
L_{m}^{\alpha}(t)\nonumber\\
\hspace{-6mm}&&=(-1)^{n+m}\frac{(m+\alpha)!}{m!}e^{-y}L_{n}^{m-n}(y)L_{m+\alpha}^{n-m}(y).
\ea
Expanding the left and right sides in $y$ we find the standard
orthogonality relation,
\ba
\int_{0}^{\infty}dt\,e^{-t}t^{\alpha}L_{n}^{\alpha}(t)
L_{m}^{\alpha}(t)=\frac{\Gamma(\alpha+n+1)}{n!}\delta_{nm}, \ea and the
integrals \ba &&\int_{0}^{\infty}dt\,e^{-t}t L_{n}(t)L_{m}(t)=
(-1)^{n+m}\left[(2n+1)\delta_{nm}\right.\nonumber\\
&&\left.+(m+1)\delta_{n,m+1}+(n+1)\delta_{m,n+1}\right],\\
&&\int_{0}^{\infty}dt\,e^{-t}t^{2} L^{1}_{n}(t)L^{1}_{m}(t)=
(-1)^{n+m}(m+1)(n+1)\nonumber\\
&&\left[2\delta_{nm}+\delta_{n,m+1}+\delta_{m,n+1}\right].
\nonumber\\
\ea
For the integrals of the functions $B_{i}(\lambda)$
 we get
$ \int_{-\infty}^{0}dE B_{0}(\lambda)=0, $ while \be
\int\limits_{-\infty}^{0}dE B_{1}(\lambda)=\frac{\hbar
v_{F}}{l}6\sqrt{2}\zeta(-1/2), \label{integral-B1} \ee where $\zeta(z)$ is the
Riemann zeta function.

\end{document}